\documentclass[letterpaper,11pt]{article}
\pdfoutput=1
\usepackage{jinstpub}
\usepackage{lineno}
\usepackage[colorlinks=true]{hyperref}
\title{Design and construction of MuTe: a hybrid Muon Telescope to study Colombian Volcanoes}

\author[a, 1]{J. Pe\~na-Rodr\'{\i}guez, \note{Corresponding author.}}
\author[a]{J. Pisco-Guabave}
\author[b]{D. Sierra-Porta}
\author[c]{M. Su\'arez-Dur\'an}
\author[d]{M. Arenas-Fl\'orez}
\author[d]{L. M. P\'erez-Archila}
\author[a]{J. D. Sanabria-G\'omez}
\author[e,f,g]{H. Asorey}
\author[a,h]{L. A. N\'u\~nez.}

\affiliation[a]{Escuela de F\'isica, Universidad Industrial de Santander,  Bucaramanga-Colombia.}
\affiliation[b]{Departamento de F\'isica, Universidad de Los Andes, Bogot\'a-Colombia;}
\affiliation[c]{Departamento de F\'{\i}sica y Geolog\'{\i}a, Universidad de Pamplona, Pamplona-Colombia}
\affiliation[d]{Escuela de Ingenier\'ia El\'ectrica, Electr\'onica y Telecomunicaciones, Universidad Industrial de Santander,  Bucaramanga-Colombia.}
\affiliation[e]{Instituto de Tecnolog\'{i}as en Detecci\'{o}n y Astropart\'{i}culas, Centro At\'{o}mico Constituyentes, Comisi\'{o}n Nacional de Energ\'{i}a At\'{o}mica, Buenos Aires-Argentina}
\affiliation[f]{Centro At\'{o}mico Bariloche, Comisi\'{o}n Nacional de Energ\'{i}a At\'{o}mica, San Carlos de Bariloche-Argentina}
\affiliation[g]{Escuela de Producci\'{o}n, Tecnolog\'{i}a y Medio Ambiente, Universidad Nacional de R\'{i}o Negro, San Carlos de Bariloche-Argentina}
\affiliation[h]{Departamento de F\'isica, Universidad de Los Andes, M\'erida-Venezuela.}

\emailAdd{jesus.pena@correo.uis.edu.co}

\abstract{We present a hybrid Muon Telescope, MuTe, designed and built for imaging active Colombian volcanoes. The MuTe has a resolution of tens of meters, low power consumption, robustness and transportability making it suitable for using in difficult access zones where active volcanoes usually are. The main feature of MuTe is the implementation of a hybrid detection technique combining two scintillation panels for particle tracking and a Water Cherenkov Detector for filtering background signals due to the electromagnetic component of extended air showers and multiple particle events. MuTe incorporates particle-identification techniques for reducing the background noise sources and a discrimination of fake events by a picosecond Time-of-Flight system.  We also describe the mechanical behavior of the MuTe during typical tremors and wind conditions at the observation place, as well as the frontend electronics design and power consumption.}


\arxivnumber{2004.09364} % only if you have one

\begin{document}
\maketitle
\flushbottom
\section{Introduction}
\label{sec:intro}

Muography or muon radiography is a non-invasive technology whose primary purpose is to obtain digital images from the density contrasts due to the different inner structures of objects by analyzing the atmospheric muon flux transmitted through them \cite{Kaiser2019, Bonomi2020, Bonechi2020}. Nowadays there are several emerging academic and commercial applications such as the detection of hidden materials in containers\,\cite{Blanpied2015}, archaeological building scanning\,\cite{Morishima2017, GomezEtal2016}, nuclear plant inspection\,\cite{Fujii2013}, nuclear waste monitoring, underground cavities\,\cite{Saracino2017}, the overburden of railway tunnels\,\cite{ThompsonEtal2019} and vulcanology applications (see, e.g.,\,\cite{TanakaOlah2019} and references therein). In Colombia, there are more than a dozen active volcanoes representing significant risks to the nearby population\,\cite{Cortes2016, Agudelo2016, Munoz2017}. This motivated local research groups to explore possible applications of the muography technique\,\cite{AsoreyEtal2017B, SierraPortaEtal2018, PenaRodriguezEtal2018, GuerreroEtal2019, ParraAvila2019, PenaRodriguez2019}.

Muons are elementary particles, two hundred times heavier than electrons and with a lifetime of approximately $2.2~\mu$s. They are produced by the interaction of particles reaching the Earth's atmosphere from galactic and extragalactic sources. These extraterrestrial impinging particles -- called cosmic rays-- generate showers of secondary particles with a significant presence of muons, produced by decaying charged pions and kaons. The energy spectrum of muons at sea level has a maximum at around $4$~GeV with a flux of $\sim 1$~cm$^{-2}$~min$^{-1}$ \cite{nakamura2010review}.

Despite a great deal of work in these areas, some particular problems exist and are still being addressed today:
\begin{itemize}

\item The low muon flux across the scanned object due to: (a) the muon flux diminishes with higher zenith angles; and (b) the flux is reduced ($\sim$two orders) in crossing a $1$~km path-length of standard rock \cite{groom2001muon, groom2000passage}. 

%\item The low muon flux across the scanned object due to: (a) the muon rate decreases about two orders of magnitude at quasi-horizontal zenith angles; and (b) the muon flux is reduced two orders more because the energy loss after traversing a $1$~km path-length of standard rock \cite{groom2001muon, groom2000passage}. 

%The low muon flux across the scanned object. The muon rate decreases about two orders of magnitude after traversing a $1$~km path-length of standard rock at quasi-horizontal zenith angles \cite{groom2001muon, groom2000passage}.

\item The background produced by charged particles: upward charged particles \cite{jourde2013experimental}, Extensive Air Showers (EAS)  \cite{nishiyama2014experimental, Olah2017ICRC, KUSAGAYA2015, Bene2013, Olh2017}, and scattered low momentum muons ($< 1$~GeV/c) \cite{nishiyama2016monte, Gomez2017, Olh2018, Olah2018Invest, ambrosino2015joint}. These particles cause an overestimation of muon flux with a corresponding underestimation of the density distribution inside the volcano \cite{carbone2013experiment, nishiyama2016monte}.
\end{itemize}

The Colombian Muon Telescope \cite{AsoreyEtal2017B, SierraPortaEtal2018} is a hybrid instrument suitable for different geophysical scenarios. MuTe employs a hodoscope manufactured of plastic scintillator bars to determine the direction of particles impinging the detector. Additionally, MuTe incorporates particle-identification techniques for reducing the background noise sources\cite{Bonechi2020, PenaRodriguez2019}. A Water Cherenkov Detector (WCD) measures the energy loss of charged particles filtering the noise due to the soft-component of EAS (electrons and positrons), and of particles arriving simultaneously. Discrimination of fake events due to scattered and backwards muons is addressed using a picosecond Time-of-Flight system.

In this paper, we report the main features of the Colombian muon telescope, which considers the signal-to-noise problems mentioned above. In section \ref{detector} we present the MuTe features, while section \ref{mechanical} describes the mechanical response of MuTe to environmental field conditions. Section \ref{daq} examines the data acquisition system, trigger mechanism and power consumption of MuTe; while in section \ref{measurement}, we present the first MuTe flux data, with an estimation of the background obtained with the WCD. Finally, in section \ref{conclusions} conclusions and final remarks are presented.

\section{The hybrid detector}
\label{detector}
In this section, we present the most significant characteristics of our muon telescope, MuTe: the event tracking and the signal-to-background discrimination. Our hybrid instrument consists of two independent detectors: a scintillator hodoscope and a WCD. A similar hybrid measurement technique has been previously implemented in the Pierre Auger Observatory to study the composition of primary cosmic rays \cite{aab2017muon, aab2016prototype}.

\subsection{Scintillator hodoscope: the tracking device}
The MuTe hodoscope consists of two detection panels, each with $60$ scintillator bars, separated from each other by a configurable distance, which is typically $\sim 250$~cm. Each plastic scintillator strip is made of a polystyrene base of Styron $665$-W doped with $1\%$ of $2.5$-diphenyloxazole (PPO) and $0.03\%$ of $1.4$-bis-[2-(5-phenyloxazolyl)]-benzene (POPOP), co-extruded with a $0.25$~mm thick high reflectivity coating of $\text{TiO}_{\text{2}}$ \cite{PlaDalmau2003}. 

\begin{figure}[htb]
\centering
\includegraphics[width=1\columnwidth]{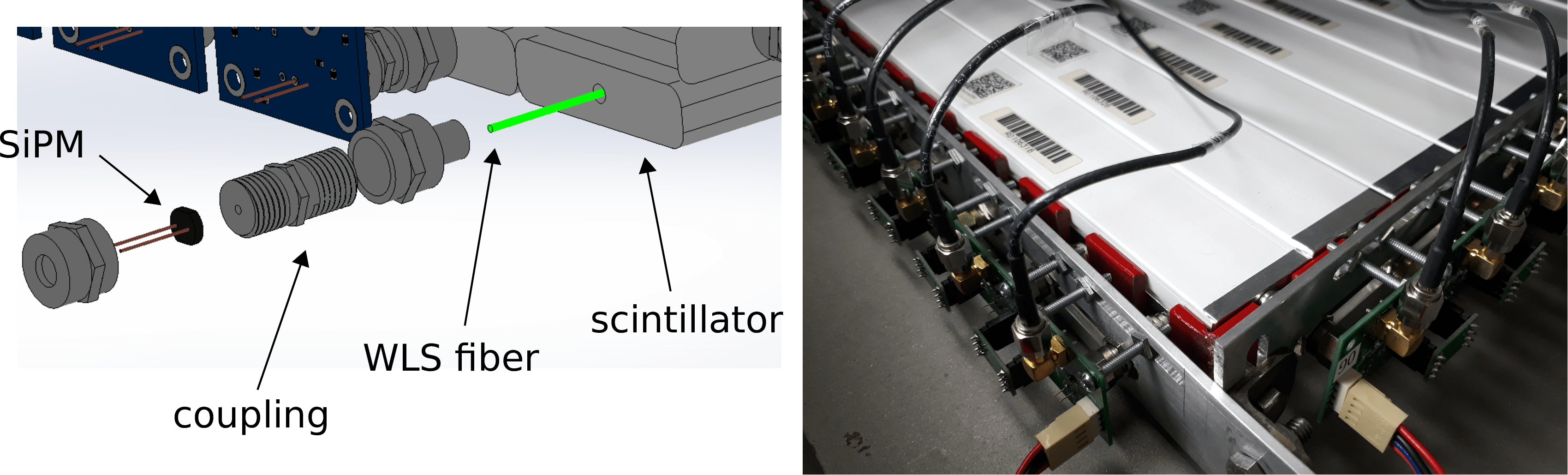}
\caption{Mounting and coupling details for one detection panel. (Left) Mechanical assembly of the scintillator bar, the Saint-Gobain BCF-92 WLS fiber and the Hamamatsu S13360-1350CS SiPM. (Right) Scintillator panel with the SiPM electronics front-end and signal transmission cables (coaxial RG-174U).}
\label{fig:frame}
\end{figure}

A $1.2$~mm diameter wavelength shifting fiber (Saint-Gobain BCF-92) is placed inside a co-extruded hole ($1.8$~mm diameter) in the scintillator strip. The fibre has a core refraction index of $1.42$, an absorption peak at $410$~nm, and an emission peak at $492$ nm. It captures the photons produced by the impinging charged particles and carries them to a Hamamatsu (S13360-1350CS) Silicon Photomultiplier (SiPM). The WLS fibre is attached to the SiPM sensitive area using a mechanical coupling (See figure \ref{fig:frame}). 

The scintillator panels are build up as an array consisting each of $30$ horizontal (X) and $30$ vertical (Y) bars, each comprising $900$ pixels of $16$~cm$^2$ active area. The two X and Y scintillator layers are mounted inside a $0.9$~mm thick stainless steel box, with the SiPM electronics, the temperature/pressure sensors (HP03), and the coaxial cables for signal transmission. A second steel housing encloses the electronics readout, the ToF system, and the power supply, keeping all the components insulated and protected from environmental conditions. 

The hodoscope reconstructs the arrival direction of muons by taking into account the pair of pixels activated in each panel. The total aperture angle and the angular resolution of the telescope varies by changing the distance between the scintillator panels. The distance between the detector and the volcanic structure as well as the separation of the panels define the spatial resolution of this telescope as
\begin{equation}
\Delta x=L\times\Delta\theta=L\times \arctan{\frac{2dD}{D^2+4d^2i(i+1)}},
\end{equation}
where $\Delta\theta$ is the angular resolution, $L$ is the distance to the target, $D$ is the separation between the panels, $d$ is pixel width, and $i$ represents the $i$th illuminated pixel. For instance, for inter-panel distances of $150$~cm, $200$~cm and $250$~cm, the total angular aperture is $1.3$~rad, $1.1$~rad and $0.9$~rad giving angular resolutions of $53$~mrad, $40$~mrad and $32$~mrad, respectively. Taking into account a distance to the volcano of $900$~m, from the previous angular resolutions we obtain spatial resolutions of $48$~m, $36$~m and $28$~m respectively.

The multiple scattering of muons in the rock and the air worsens the effective spatial resolution of the telescope. The maximum scattering angle estimated is $\sim~1.5^{\circ}$ taking into account the minimum energy needed by muons to cross standard rock path-lengths between $10$~m and $1000$~m. Nevertheless, muons must have extra energy to reach the detector after leaving the scanned object. Those emerging from the scanned object with an energy $>1$~GeV have a scattering angle $\sim~1^{\circ}$ \cite{Suarez2019}, generating a blurring of $\sim$ $31.4$~m for $L=900$~m. The exposure time of MuTe to achieve a sensible contrast is $\sim$100 days \cite{AsoreyEtal2017B}.

\subsection{Telescope acceptance}
%\label{acceptan}
The acceptance of the instrument affects the measured particle flux depending on the telescope's geometric parameters: number of pixels in the panel ($N_x~\times~N_y$), pixel size ($d$) and inter-panel distance ($D$). The number of detected muons $N(\varrho)$ \cite{LesparreEtal2010}, can be defined as
\begin{equation}
N(\varrho)=\Delta T~\times~\mathcal{T}\times I(\varrho), \label{Nmuons}
\end{equation}
where $I(\varrho)$ is the integrated flux (measured in cm$^{-2}$~sr$^{-1}$~s$^{-1}$), $\mathcal{T}$ the acceptance function (measured in cm$^{2}$~sr), and $\Delta T$ the recording time. The integrated flux depends on the opacity ($\varrho$): the amount of matter crossed by the muons.

\begin{figure}[htb]
\centering
\includegraphics[width=0.48\columnwidth]{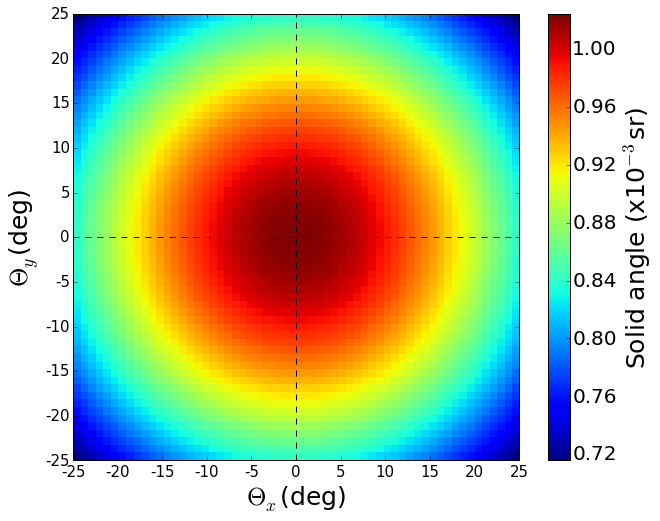}
\includegraphics[width=0.48\columnwidth]{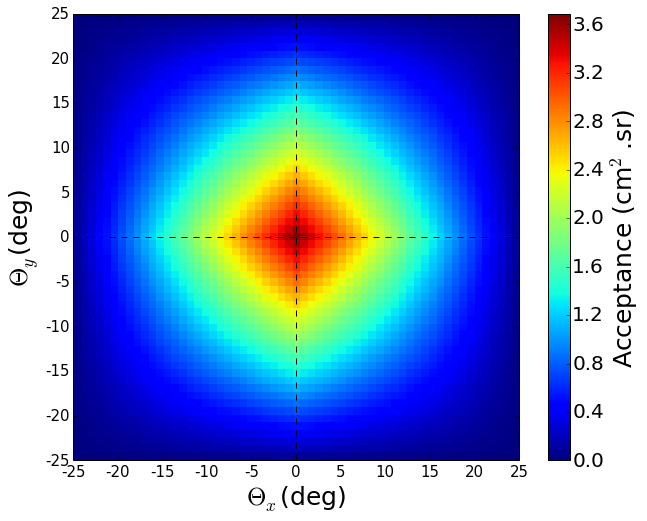}
\caption{Angular resolution (left), and acceptance function (right) for the MuTe hodoscope with $N_x=N_y=30$, $d=4$~cm and $D=250$~cm versus incoming direction. The maximum solid angle is $1.024~\times~10^{-3}$~sr for perpendicular trajectories where the acceptance rises up to $3.69$~cm$^{2}$~sr.}
\label{fig:acceptance}
\end{figure}

For a particular trajectory $r_{m,n}$ displayed by a pair of illuminated pixels on both panels, one can calculate the solid angle $\delta\Omega(r_{m,n})$ and the detection area $S(r_{m,n})$. All pairs of pixels with the same relative position, {$m=i-k$, $n=j-l$}, share the same direction, $r_{m,n}$ and the same $\delta\Omega(r_{m,n})$. This means directions normal to the hodoscope plane, have the larger detection area, while directions crossing corner-to-corner have a smaller solid angle and detection surface. The acceptance is obtained \cite{LesparreEtal2010} multiplying the detection area by the angular resolution,
\begin{equation}
\mathcal{T}(r_{m,n})=S(r_{m,n})\times \delta\Omega(r_{m,n}).
\end{equation}

A hodoscope with two matrices of $N_x\times N_y$ pixels has $(2N_x-1)\times(2N_y-1)$ discrete directions $r_{m,n}$, spanning an solid angle $\Omega$. Our telescope equipped with $900$ pixels in each panel is able to reconstruct $3481$ discrete directions. In figure \ref{fig:acceptance}, we show the angular resolution and acceptance function for the Mute hodoscope with $N_x=N_y=30$ scintillator bars, pixel size ($d=4$~cm) and $D=250$~cm. The total angular aperture of the telescope with that configuration is roughly 50$^{\circ}$(0.9 rad) with a maximum solid angle of $1.024\times 10^{-3}$~sr at $r_{0,0}$ corresponding to the largest acceptance of $\approx 3.69$~cm$^{2}$~sr.

The MuTe acceptance may be compared with other muon telescopes mentioned in the literature: a) $N_x=N_y=12$, $d=7$~cm and $D=100$~cm; with an acceptance $\mathcal{T}=30$~cm$^{2}$~sr and angular resolution $< 0.018$~sr \cite{UchidaTanakaTanaka2009space}. b) $N_x=N_y=16$, $d=5$~cm and $D=80$~cm; with a acceptance $\mathcal{T}=25$~cm$^{2}$~sr and angular resolution $< 0.015$~sr \cite{LesparreEtal2010}.

\subsection{Time-of-Flight: momentum measurement}
Time-of-Flight methods have been applied in muography to distinguish backward moving particles from the incident ones \cite{jourde2013experimental}. Particles entering from the rear side of the detector represent roughly $44\%$ of background noise for zenith angles above $81^{\circ}$ \cite{nishiyama2016monte}.

MuTe performs ToF measurements for identifying backward particles as well as low momentum ($< 1$~GeV/c) muons which are scattered by the volcano contributing to the background noise. The MuTe ToF system was implemented on a Field Programming Gate Array (FPGA) employing a Time-to-Digital Converter with a time resolution of $\sim 40$~ps, which measures the time-lapse of the crossing particles between the front and rear panel.

% \begin{figure}[htb]
% \centering
% \includegraphics[width=0.5\columnwidth]{Figures/ToF_Resolution_250cm.eps}
% \caption{ToF resolution required to estimate the muon momentum to within $\pm 10$~MeV/c (black line) or $\pm 100$~MeV/c (red line), for an inter-panel distance of $250$~cm. The blue line represents the up-to-date MuTe ToF resolution ($40$~ps).}
% \label{fig:ToF_Resolution}
% \end{figure}

Taking into account the ToF $t$ of particles crossing the hodoscope in a given trajectory $d$, and the particle identification provided by the WCD (distinguishing muons from electron/positrons) we can estimate particle momentum as follows
\begin{equation}
p = \frac{m_0 c d}{\sqrt{c^2t^2-d^2}} \, ,
\end{equation}
where $m_0$ is the rest mass of the charged particle ($ 105.65$~MeV/c$^2$ for muons and $0.51$~MeV/c$^2$ for electrons/positrons) and $c$ the speed of light. 

The uncertainty in the momentum estimation depends on the error of the ToF measurement and the error of the trajectory length, as
\begin{equation}
\sigma_p^2 = \left( \frac{\partial p}{\partial t} \right)^2 + \left( \frac{\partial p}{\partial d} \right)^2
\end{equation}

Measuring the momentum allows us to set a threshold of $1$~GeV/c, above which the influence of noise due to soft muons is negligible \cite{nishiyama2016monte, nishiyama2014experimental, Olh2018, Olh2017, ambrosino2015joint}. In order to establish such a cutoff, we calculate the ToF resolution requirements, which depends on the momentum resolution we want to obtain. For perpendicular tracks to the hodoscope plane, we need a ToF resolution of $10$~ps to differentiate muons with momentum $>~1$~GeV/c with an error of $\pm~0.1$~GeV/c. 

%In figure \ref{fig:ToF_Resolution} we show the ToF resolution $\delta_t$ versus the particle momentum for estimated errors of $\pm~10$~MeV/c and $\pm~100$~MeV/c and for an inter-panel distance of $2.5$~m. 

The ToF measurements must be compensate for the signal delay depending on the length of the transmission lines and the impinging point of the particle in the scintillator panels. We found the signal is delayed $5.05$~ns/m in the transmission lines and $77$~ps/cm in the scintillator bars.

The estimated ToF resolution of MuTe is $\sim138$~ps establishing a momentum threshold from $0.4\pm~0.1$~GeV/c. The pixel-related error is $\sim89$~ps taking into account the pixel spatial resolution ($4$~cm/$\sqrt{12}$) and the scintillator bar delay ($77$~ps/cm). This value increases to $\sim97$~ps by adding in quadrature the electronics resolution ($40$~ps). The effective coincidence timing resolution for the hodoscope is the panel time resolution multiplied by $\sqrt{2}$ \cite{Moses2010}. 

\subsection{Water Cherenkov Detector: deposited energy measurement}

Water Cherenkov Detectors, widely used in cosmic-ray observatories, have high acceptance, reasonable efficiency, and $\sim 100\%$ duty cycle. These devices, implemented with few cubic meters of water and with one or more photomultiplier tubes (PMTs), record the Cherenkov radiation produced by charged crossing particles moving with a velocity greater than the speed of light in water. They are sensitive to the muonic and electromagnetic component of air showers \cite{Auger2015}, and also detect --indirectly-- high energy photons by pair production ($\gamma \rightarrow$ e$^{\pm}$) \cite{allard2007detecting, allard2008use, allekotte2008surface}. 

The MuTe's WCD is a $3.2$~mm thick stainless steel cube of $1.2$~m sides, coated inside with Tyvek diffuser sheets, which enhance the reflectivity for the Cherenkov photons. An eight inch PMT (Hamamatsu R5912) with a quantum efficiency of $22\%$ at $390$~nm-- acts as the photosensitive device. The number of photons detected by the PMT can be associated with the energy deposited by the crossing particle, allowing us to differentiate between muons and the electromagnetic component of EAS (photons, electrons, and positrons) \cite{Billoir2014}. The EM component is one of the most important noise sources in muography \cite{KUSAGAYA2015, Nishiyama2014Noise, Marteau2012Noise}. At ground level, the most probable muons ($\sim 4$~GeV) can traverse the whole WCD losing up to $240$~MeV ($2$~MeV/cm along $120$~cm) for perpendicular trajectories; while the most probable electrons ($\sim 20$~MeV) stop in $10$~cm of water losing $2$~MeV/cm  \cite{groom2001muon,groom2000passage,lohmann1985energy,olive2014passage,Vasquez2018, Motta2018}.

The WCD detects charged particles coming from all directions due to its $2\pi$ acceptance with a deposited energy resolution of $\sim 0.72$~MeV and a measuring range from $50$~MeV (which is the typical electronic noise) up to $\sim 1.5$~GeV (which is the largest value that does not saturate the electronics readout).This property allows the MuTe to monitor the local variations of the secondary particle flux over time \cite{Leon2018} and also, to distinguish particles coming from the volcano direction by coincidence with the hodoscope trigger.

\section{Mechanical response}
\label{mechanical}
\subsection{Structural design}
\begin{figure}[htb]
\centering
\includegraphics[width=1\columnwidth]{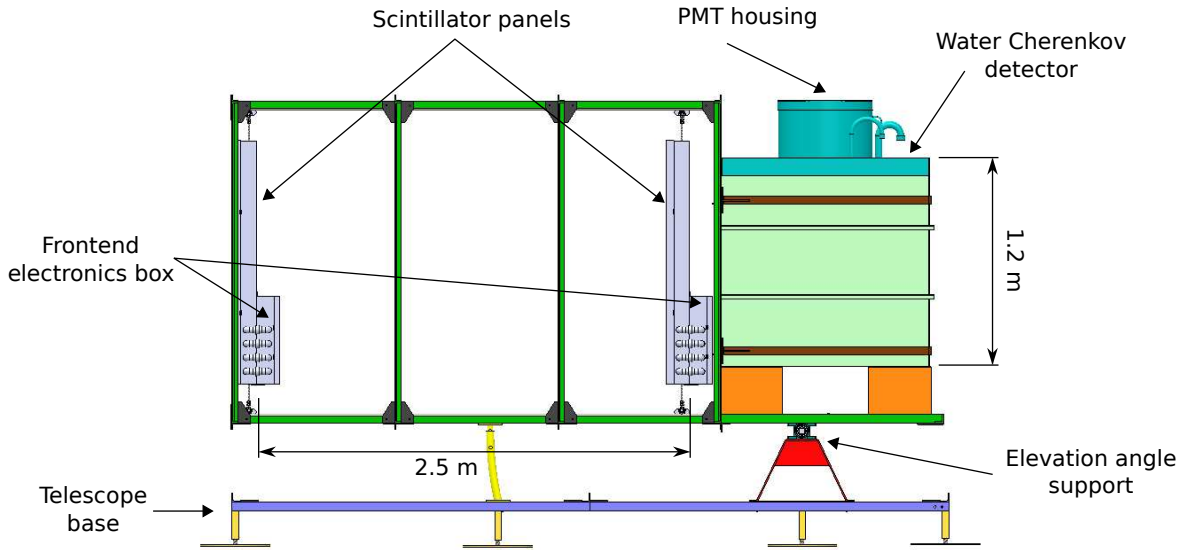}
\caption{MuTe lateral view. The WCD, placed on the centre of mass of the structure, benefits the angular elevation (maximum $15^{\circ}$, in $3^{\circ}$ steps). The inter-panel distance can vary from  $40$~cm up to $250$~cm. The electronics front-end boxes and the PMT housing are isolated from rain and humidity.}
  \label{fig:Structure}
\end{figure}

As shown in figure \ref{fig:Structure}, the hodoscope, the WCD, the electronic readout, and the central monitoring server are all mounted on a sturdy metallic structure. The frame consists of a $4.2$~m~$\times~2.8$~m~$\times~1.8$~m parallelepiped-shaped structure constructed of steel angles ASTM A-36 of $3.2$~mm thickness and mechanically attached with screws ($0.5$~inch diameter). The telescope can be raised up to $15^{\circ}$ with respect to the horizontal. 

%Notice that the rear panel is fixed to one of the walls of the WCD, while the frontal panel can slide on a rail of $2.8$~m length, providing a significant variable angular resolution for the telescope. 

 It is necessary to study the MuTe's mechanical behaviour in the volcanic environment. In the following section, we analyze the stress load and the vibration of the instrument, by taking into account it's structural design as well as the mechanical stresses due to tremors and wind conditions. Such simulations, using Finite Element Analysis (FEA), predict the physical behaviour of the instrument with linear/non-linear, and static/dynamic analysis capabilities. We employed the \textsc{Solidworks 3D CAD Modeling Software} with the package \textsc{Solidworks Simulation} for the structural analysis.

\subsection{Vibration analysis and tremor response}
We calculated the natural frequencies and vibration patterns of the instrument under the external influences of wind sources and typical volcano seismic activity which could affect the integrity of the telescope. 

Volcanic tremors and internal movements can be distinguished\cite{mcnutt1992volcanic,londono2001spectral,langer2006automatic,chouet2003volcano}: 
\begin{itemize}
    \item volcano-tectonic earthquakes associated with fracturing that occur in response to stress changes in the active areas due to fluid movements with frequency peaks between $2$ and $15$~Hz.
    \item long period tremors with frequencies of $1$ to $2$~Hz, attributed to pressure changes in cracks, cavities and ducts. 
\end{itemize}
\begin{table}[htb]
\begin{center}
\begin{tabular}{ccccc}
\hline
& {\bf Frequency}  & {\bf Max. vibration}\\
{\bf Mode}  & {\bf (Hertz)}  & {\bf (Hertz)} \\
\hline
1 & 1.6 & 0.01272\\ 
2  & 5.0149  & 0.0113 \\
3  & 5.6445  & 0.0303 \\
4  & 7.5633  & 0.0361 \\
5  & 7.5702  & 0.0166 \\
\hline
\end{tabular}
\end{center}
\caption{Vibration analysis of the instrument. The first column indicates the natural frequencies of the MuTe structure., while the second one shows the maximum vibration reaction when the structure is under resonance.}
\label{Table_nat_frec1}
\end{table}

As can be seen in table \ref{Table_nat_frec1}, the instrument undergoes negligible mechanical affectation due to displacements caused by tremors or other manifestations, inherent to volcanic environments having a frequency range from $1.6$~Hz to $7.5$~Hz. The structure's reaction (vibration frequencies not exceeding $0.04$~Hz) guarantees its structural integrity against seismic events triggered by volcanic activity. 

\subsection{Static and wind load}
The aim of the static load analysis is to simulate the behaviour of the instrument against deformations that may lead to structure failure.

\begin{figure}[htb]
\centering
\includegraphics[width=0.48\columnwidth]{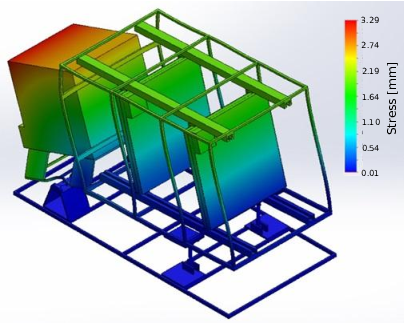}
\includegraphics[width=0.48\columnwidth]{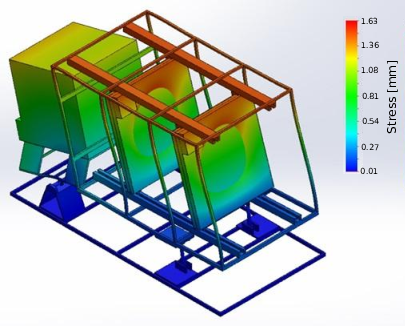}
\caption{Left and right plates illustrate the stress graph resulting from the static and dynamic --wind action-- load analysis, respectively. The maximum material deflection is about $3.29$~mm in the rear part of the WCD, which is under high pressure due to the water weight.  The maximum wind pressure occurs in the front of the scintillator panels suffering a mechanical displacement about $1.63$~mm. To determine dynamical wind loads on the instrument, we used meteorological data from IDEAM. The maximum wind speed reported is $30$~m/s, with an occurrence probability of $4\%$}
\label{fig:stress}
\end{figure}

The MuTe structure is mostly ASTM A-36 steel. The possible primary loads for the structure arise from two sources: the water volume inside the WCD ($\sim 1728$~Kg) and the metal frames for the scintillator panels ($\sim 70$~Kg each). The simulation mesh was $2.6~\times~10^6$ finite elements with $15 \pm 5$~mm size. Figure \ref{fig:stress} (left) displays the simulation results with displacements ranging from  $0$~mm up to $3.29$~mm with the maximum peak stress under the WCD. However, such deformations do not represent any considerable mechanical problem for the instrument.

To determine dynamical wind loads on the instrument, we use meteorological data from IDEAM and the maximum wind speed reported is $30$~m/s, with an occurrence probability of $4\%$. The right panel in figure \ref{fig:stress} illustrates the structure stress due to wind load. The mechanical structure suffers displacements up to $1.63$~mm in the frontal part of the scintillator panel. However, this displacement is negligible, and the instrument will not experience significant deformations.

\subsection{Heat dissipation in the structure}
We simulated the temperature distribution based on the thermal inputs (heat loads) and outputs (heat losses) by considering the conduction, convection and thermal irradiation due to the detector environment. Such processes include environmental temperature, solar radiation, wind cooling, and the heat dissipated by the electronics.

This thermal analysis allowed us to understand the heat transfer along the structure and how it may affect the detector components. Instrument safety and reliability in the field is an essential factor to consider since the characteristics of many components (SiPMs, scintillation bars, etc.) depend on temperature. 

We performed the thermal structure analysis using the heat module of \textsc{Solidworks Modeling Software} with the parameters shown in Table \ref{instr_mat}. Again, IDEAM provided the average temperature, radiation and wind speed on the observation place.

\begin{figure}[htb]
\centering
\includegraphics[width=1\columnwidth]{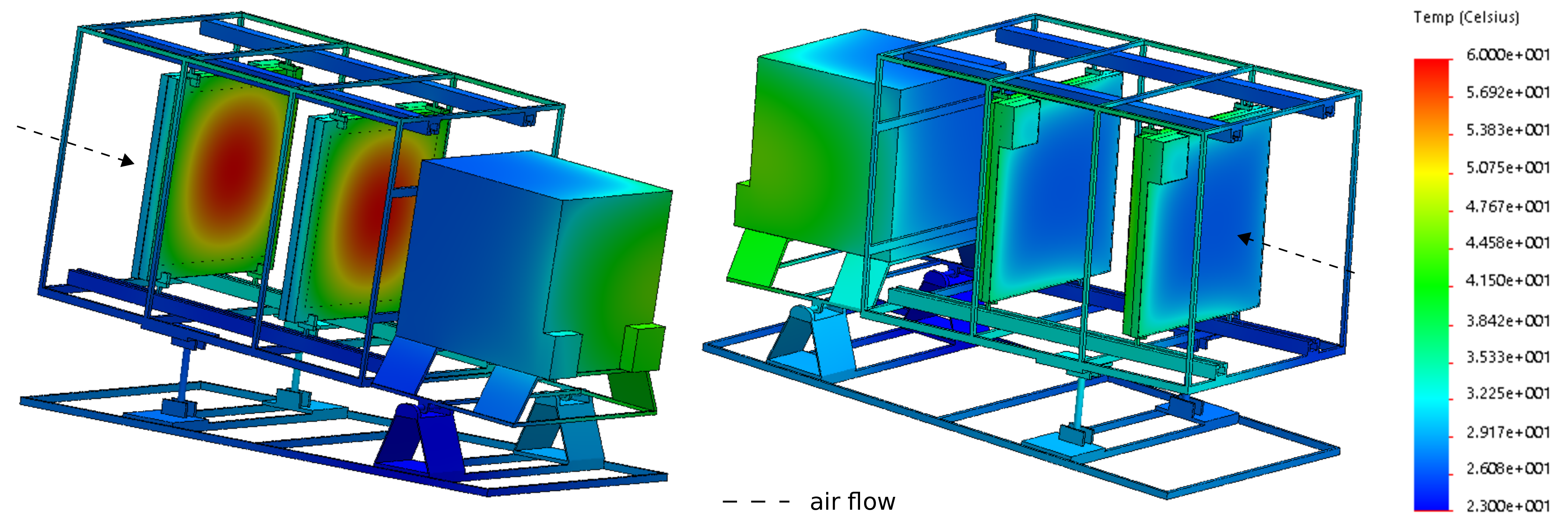}
\caption{Temperature distribution in the MuTe. The maximum heat area ($60^{\circ}$C) is on the rear side of the scintillator panels due to the solar radiation, while the front sides are cooled by the wind convection. The WCD has its heat-dissipation mechanism due to its water volume content.}
\label{fig:temp_graph}
\end{figure}

In figure \ref{fig:temp_graph} we see that the areas of maximum temperature in the detector, reaching $\sim 60^{\circ}$C, at the centre of the scintillation panels where the solar radiation heats a large surface, and an average of $23^{\circ}$C in the remaining structure. The WCD is a good heat dissipator due to its large metallic area and water content ($\sim 1.7$~m$^3$) attaining a maximum temperature of $40^{\circ}$C; while the front side of both panels have a lower temperature than the rear side since the wind flow generates a cooling process by convection. 

\begin{table}
\begin{center}
\begin{tabular}{ll|ll}
\hline
\multicolumn{2}{c}{\bf DETECTOR MATERIALS} & \multicolumn{2}{|c}{\bf HEAD SOURCES DATA}\\
\hline
Structure Material: & AISI 1020 & Sky temperature & -10 $^{\circ}$C \\
Model type: & Linear elastic isotropic & Electronic box WCD & 5.2 W \\
Thermal conductivity: & 47 W/(m K) & Gen. electronic box & 12.5 W \\
Specific heat: & 420 J/(kg K) & Electronic box Scint. & 12.3 W \\
Density: & 7900 kg/m$^3$ & Sun radiation & 4500 Wh m$^{-2}$ day$^{-1}$ \\
Cherenkov medium: & Water & Convection coeficient & 10 W/(m$^2$ K) \\
Model type: & Linear elastic isotropic & Mean enviroment temp. & 16 $^{\circ}$C \\
Thermal conductivity: & 0.61 W/(m K) & Base water temperature & 10 $^{\circ}$C \\
Specific heat: & 4200 J/(kg K) & & \\
Density: & 1000 kg/m$^3$ & & \\ 
\hline
\end{tabular}
\end{center}
\caption{Instrument materials and data used in the MuTe thermal analysis implemented by using the heat module of \textsc{Solidworks Modeling Software}. The simulation input consists of the heat transfer properties of the MuTe metallic structure, as well as, the heat sources surrounding the instrument which have environmental origin or  caused by the electronics functioning. }
\label{instr_mat}
\end{table}

The temperature distribution allows us to identify the critical heat areas in the detector structure. Consequently, we can enhance heat dissipation, by shading the instrument from solar radiation as well as allowing air circulation for cooling by convection. 

\section{Electronics readout}
\label{daq}
The MuTe electronics has two main --independent but synchronized-- readout systems: one for the hodoscope and one for the WCD. 

\begin{figure}[htb]
\centering
\includegraphics[width=0.8\columnwidth]{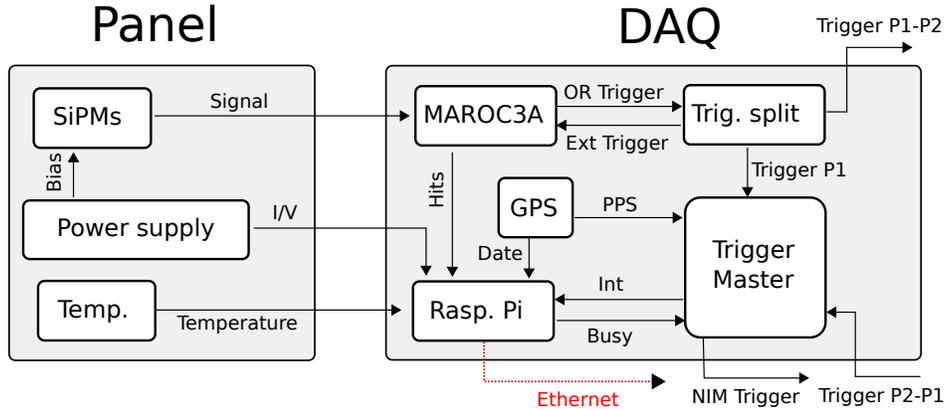}
\caption{Diagram of a single scintillator panel. Signals from the SiPMs are read out by the MAROC3 board whose slow control parameters are handled by a Raspberry Pi 2. All the detected events are time-stamped and sent via ethernet to the central monitoring server. The master trigger measures the ToF of the crossing particle and notifies the event truthfulness.}
  \label{fig:scintillatordetector1}
\end{figure}

In the hodoscope, $120$~SiPMs Hamamatsu S13360-1350CS, with a gain of $\sim 10^6$ and a photo-detection efficiency of $40\%$ at $450$~nm-- detect the light signals coming from the scintillator bars. Each SiPM has a pre-conditioning electronics for amplifying ($\times 92$) and enhancing the signal-to-noise ratio before the transmission. 

A multi-channel ASIC MAROC3 from Omega discriminates the $60$ 
signals after making a gain adjustment to reduce the bar response variability. An FPGA Cyclone III sets the MAROC3 slow control parameters (channel gains and discrimination thresholds) from Altera. We set a discrimination threshold of $8$ photo-electrons taking into account previous analysis of dark count, cross-talk and after-pulse of the SiPM S13360-1350CS \cite{Villafrades2020}.

The SCB Raspberry Pi 2 records the data from the scintillation panels when a coincidence condition is fulfilled (See section \ref{trigger}). Environmental data (temperature, barometric pressure, and power consumption) are also recorded for post-processing, status monitoring, and calibration procedures. On the other hand, the SBC controls the SiPMs bias voltage depending on the temperature via the programmable power supply C11204. The recorded events are individually time-stamped with a resolution of $10$~ns and synchronized using the PPS (Pulse Per Second) signal from a Venus GPS. A general diagram of the electronics readout for a single scintillator panel is shown in figure \ref{fig:scintillatordetector1}.

\begin{figure}[htb]
\centering
\includegraphics[width=0.9\columnwidth]{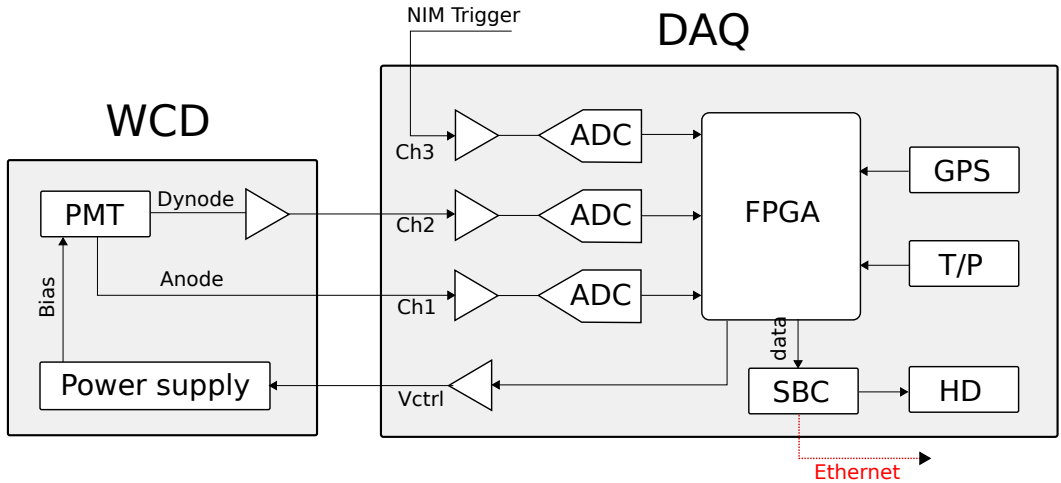}
\caption{The diagram of the WCD DAQ system. The PMT and the bias electronics are inside the WCD. A 10 bits ADC digitizes signals from the PMT anode and last-dynode and stored in a hard disk join with temperature and barometric pressure data. An FPGA sets the acquisition parameters and the event time-stamp}
  \label{fig:WCD}
\end{figure}

In the WCD, a PMT R5912 detects the Cherenkov light from the charged particles crossing the water volume. The PMT is biased through a tapered resistive chain by a high-voltage power supply EMCO C20 spanning $0$ to $2000$~V. The pulses from the anode and the last dynode --amplified $20$ times-- are independently digitized by two $10$ bits ADCs with a sampling frequency of $40$~MHz. A $12-$sample vector stores the pulse shape in each channel,  when the signal amplitude exceeds the discrimination threshold ($\sim 100$ ADC bins).  Then, a temporal label with $25$~ns resolution concatenates the event information. The timestamp is synchronized with the PPS signal from a GPS Motorola OnCore. Temperature and barometric pressure data are also recorded for the off-line analysis and data correction. An FPGA Nexys II handles the tasks of thresholding, base-line correction, temporal labelling, and temperature-pressure recording.

A third ADC channel digitizes a NIM signal coming from the hodoscope when an in-coincidence event occurs (See \ref{trigger}). The acquisition parameters of the WCD DAQ system (shown in figure \ref{fig:WCD} with discrimination thresholds and the PMT bias voltage) are set by an SCB Cubieboard 2. All the data is stored locally on an external hard disk. 

A local server collects data from the WCD and the hodoscope every $12$~hours for carrying out an \textit{in-situ} analysis. The results are sent via the GSM network to a remote server, which updates the MuTe status on a monitoring web page.

\begin{figure}
\centering
\includegraphics[width=0.8\columnwidth]{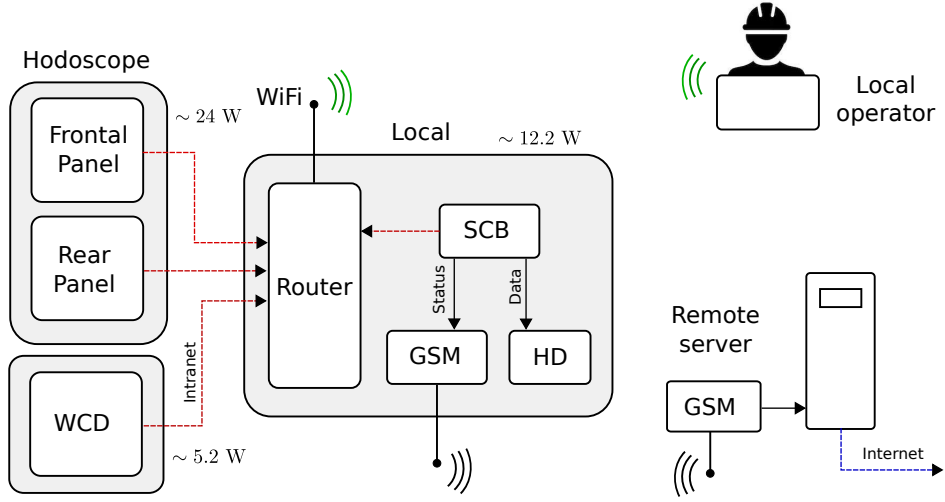}
\caption{General diagram of MuTe. The local server manages the data coming from the hodoscope and the WCD by intranet while a hard disk stores all the obtained information. Then, MuTe sends its operational status via GSM towards a remote server, and a WiFi connection link for local testing. The whole detector consumes $41.4$~W being the local server the major power load, dissipating ($\sim 12$~W) due to the operation of the hard disk, the intranet router, and the GSM transceiver.}
\label{fig:power}
\end{figure}

An intranet system connects the local server, the hodoscope, and the WCD, as shown in figure \ref{fig:power}. Additionally, the MuTe enables a wireless access point for working locally from a laptop or any other mobile device.

\subsection{Triggering system}
\label{trigger}
The MuTe triggering system determines event coincidence between the hodoscope and the WCD \cite{PenaRodriguez2019}. The Trigger $T1$ is individually enabled for rear and front hodoscope panels when the pulse amplitude exceeds the discrimination threshold value. This trigger signal splits into three sublevels: the Trigger P1-P2 for cross-checking events in-coincidence and ToF measurements, the Trigger P1 for starting the data transmission from the MAROC 3A to the SCB, and the Ext-Trigger for holding the information inside the MAROC 3A while it is read.

To identify the position of the activated pixel, MuTe counts only the events activating a vertical and a horizontal bar per panel, called trigger T2. Coincident events between the front and the rear panel in a time window ($7$ to $12$~ns) classify as crossing particles, determine the trigger T3, and estimate the particle flux across the hodoscope. The coincidence window takes into account the time needed by a particle travelling at the speed of light through two paths: the shortest ($2.5$~m) and the longest ($3.5$~m).

When the WCD detects a particle, the trigger T4 is activated. Next, trigger T5 determines coincidence of the events between the hodoscope and the WCD; this trigger is also called hybrid trigger (T5 = T3 \textbf{AND} T4). The NIM Trigger signal is digitized by the third ADC channel of the WCD for labelling the events in-coincidence with the hodoscope. All the time delays due to the transmission of the signals are considered for data analysis.

\subsection{Power consumption and operating autonomy}
Electrical power independence is a crucial parameter for our MuTe detector because it has to operate autonomously on a distant location. We designed a photovoltaic system taking into account the power requirement of all detector components. MuTe power supply has four photovoltaic panels of $100$W ($18$~V, $5.56$~A). This panel array provides the instrument with six days of continuous operation, which is the maximum number of consecutive cloudy days that occurred in the last $22$ years\footnote{We use meteorological information (irradiance, temperature, and cloudiness) from NASA satellites \url{https://eosweb.larc.nasa.gov/} and from the Colombia Meteorology and Hydrology National Institute, i.e. in Spanish \textit{Instituto de Hidrología, Meteorología y Estudios Ambientales} (IDEAM) \url{http://atlas.ideam.gov.co}}. 

Appendix \ref{ApA} details the estimation of the power capacity and autonomy of the Colombian Muon Telescope and figure \ref{fig:power} displays the power consumption data: $\sim 24$~W for the hodoscope,  $\sim 5.2$~W for the WCD  and $\sim 12.2$~W for the central monitoring server. The two hard disks  --used to store $470$~MB per hour of data from the hodoscope and the WCD-- are the devices with greater power consumption, but they provide almost six months of data-storage autonomy.

\section{First measurements}
\label{measurement}

In the first measurements, the MuTe hodoscope was operated in the vertical direction recording the muon flux during $15$~hours. The average counting rate was $\sim~836.3$~event/h with a discrimination threshold of $8$~photo-electrons (MIP $\sim~16$~pe). The inter-panel distance was $134$~cm, the angular aperture $82^{\circ}$ and the maximum acceptance of $12.83$~cm$^{2}$~sr. To reconstruct the particle trajectories and the flux crossing through the hodoscope, we apply a four-bar activation condition: a pair XY in the front panel and a pair XY in the rear one. 

\begin{figure}[htb]
\centering
\includegraphics[width=0.48\columnwidth]{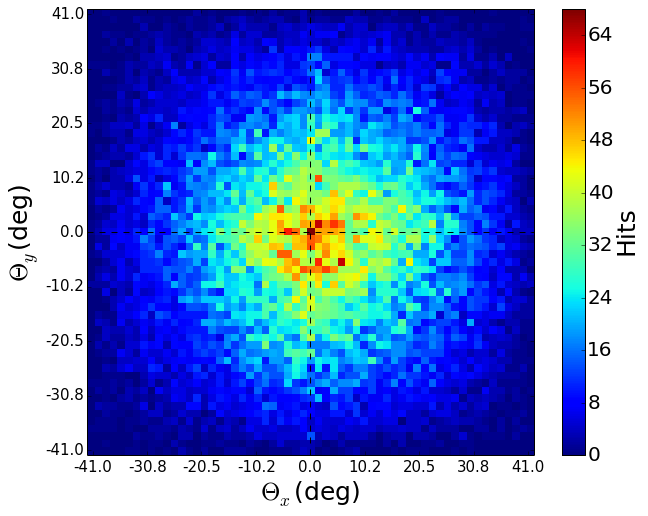}
\includegraphics[width=0.49\columnwidth]{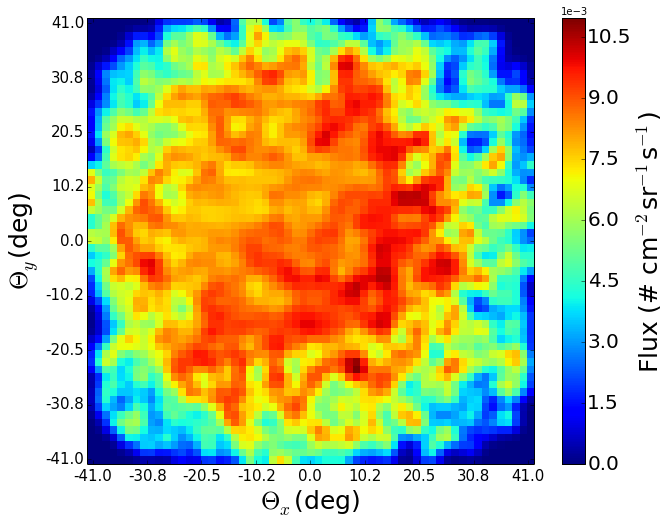}
\caption{Particle count data recorded by the hodoscope operating in the vertical direction during $15$~hours with a separation of $134$~cm between panels. The vertical flux was $\sim 10.7~\times~10^{-3}$~cm$^{-2}$~sr$^{-1}$~s$^{-1}$. As expected, the flux decreases while the zenith angle increases: for a zenith angle of $41^{\circ}$ the flux is around $4.5~\times~10^{-3}$~cm$^{-2}$~sr$^{-1}$~s$^{-1}$, a half order of magnitude lower than the flux maximum.}
\label{fig:hits_15}
\end{figure}

In figure \ref{fig:hits_15}, we display the number of hits and the particle flux recorded by the hodoscope. The maximum count was $\sim$~67 for straight trajectories ($\theta_x=\theta_y=0^{\circ}$). The number of counts decreases for non-perpendicular trajectories due to the hodoscope acceptance and the muon flux, which is modulated by the zenith angle ($\cos^2 \theta$). The estimated flux reaches a maximum of $10.7~\times~10^{-3}$~cm$^{-2}$sr$^{-1}$s$^{-1}$ which is comparable with the flux of $9~\times~10^{-3}$~cm$^{-2}$~sr$^{-1}$~s$^{-1}$, reported in reference \cite{Lesparre2012} . The variance in the flux histogram can be reduced by increasing the acquisition time.
 
\begin{figure}[htb]
\centering
\includegraphics[width=0.7\columnwidth]{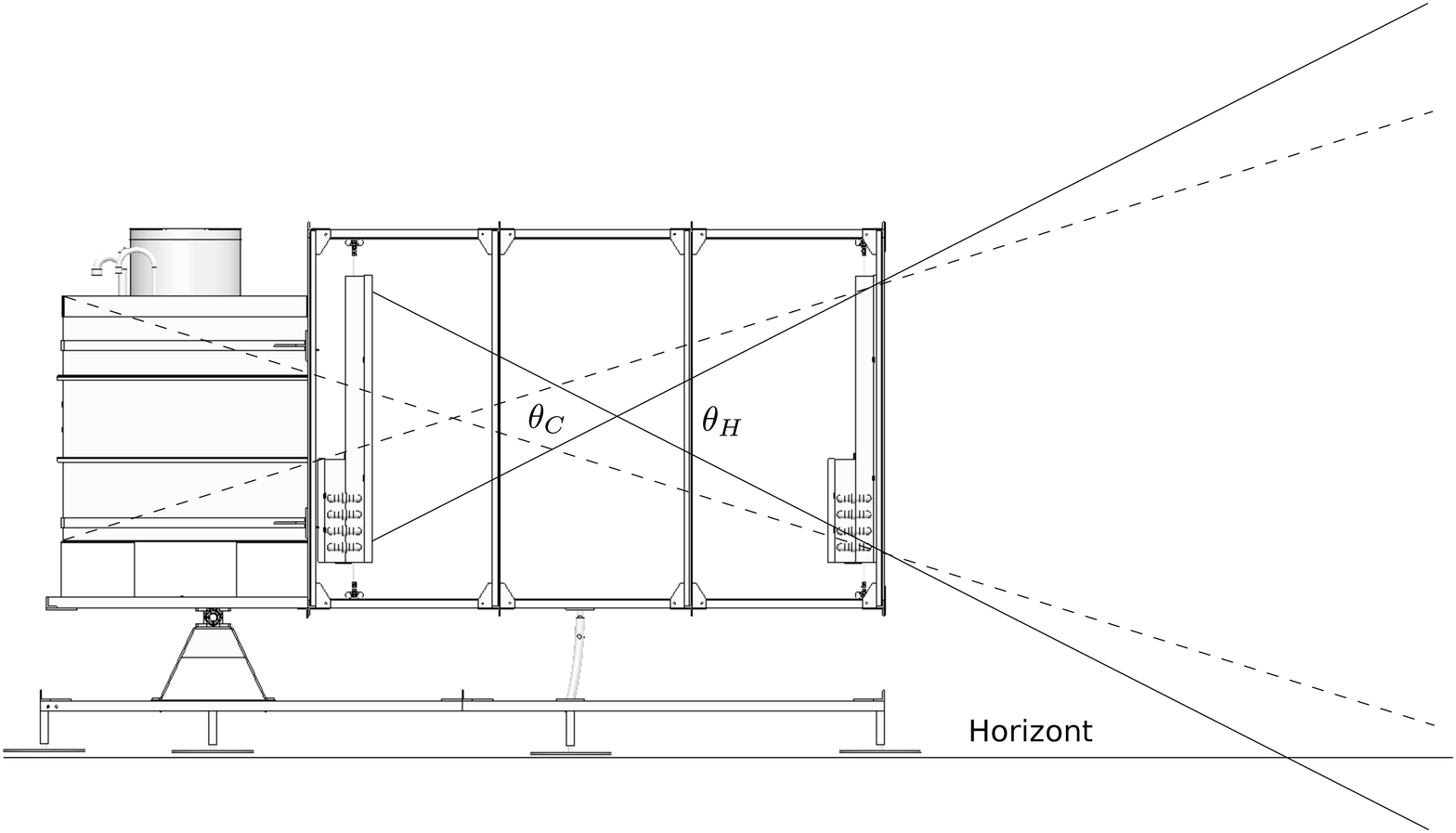}
\caption{The MuTe setup for the first field measurements. The detector is pointing towards the horizon with an elevation angle of $0^{\circ}$. The aperture of the hodoscope $\theta_H$ is $50^{\circ}$ for a separation distance between panels of $250$~cm. The aperture of the whole detector (WCD + hodoscope) $\theta_C$ is roughly $32^{\circ}$.}
\label{fig:WCDHod}
\end{figure}

\begin{figure}
\centering
\includegraphics[width=0.7\columnwidth]{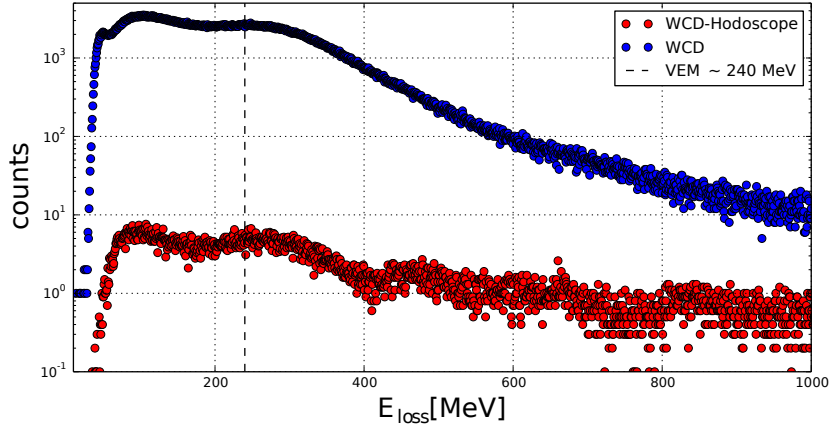}
\caption{Energy deposited in the WCD for the omnidirectional events (blue) and for the in-coincidence events between both WCD and the hodoscope (red). The dashed line represents the deposited energy of the vertical muons (VEM) which is estimated to be $240$~MeV taking into account muon losses ($\sim 2$~MeV/cm in water). The first hump corresponds to the energy deposited by electrons, positrons and gammas while above $400$~MeV events correspond to multiple particles.}
\label{fig:WCDHod_rate}
\end{figure}

Later, the MuTe was set outdoors pointing in the horizontal direction (0$^{\circ}$ elevation) as shown in figure \ref{fig:WCDHod}. The WCD and the hodoscope each detect individually but synchronized in time. The in-coincidence flux between both is two orders of magnitude lower than the events recorded by the WCD (see figure \ref{fig:WCDHod_rate}) representing only 2$\%$. This reduction in flux is due, mainly, to:
a two order of magnitude decrease in the muon  flux between the maximum at 0° zenith and at quasi-horizontal angles;  and, furthermore, the angular acceptance of the WCD is roughly 2$\pi$ while that of the hodoscope is only a fraction due to its geometry.

%the muon flux is almost two orders of magnitude weaker at quasi-horizontal angles than the maximum flux at 0${\circ}$ zenith. Furthermore, the angular acceptance of the WCD is roughly 2$\pi$ while the acceptance of the hodoscope is only a fraction of that, constrained by coincidence between both sensitive panels.

The energy deposited in the WCD (blue) and for the in-coincidence events between both WCD and hodoscope
figure \ref{fig:WCDHod_rate}, emerge from three main sources: muons, electron/positrons and multiple particle events. The muonic component represents roughly $33.6\%$ of the events ($180$~MeV$<~E_{loss}~<~400$~MeV), the electromagnetic $36$\% ($E_{loss}~<~180$~MeV), and the multiple particle $30.4\%$ ($E_{loss}~>~400$~MeV) of the histogram. 

These results show that the background (electromagnetic and multiple particles) is comparable to the signal, even greater taking into account that soft muons have not been extracted from the muonic component. On the other hand, multiple particle background made up by several particles temporally correlated, e.g. inclined cosmic showers impacting the detector \cite{Bonechi2019}, become more significant comparable in magnitude to the electromagnetic and muonic humps.

\begin{figure}[htb]
\centering
\includegraphics[width=0.8\columnwidth]{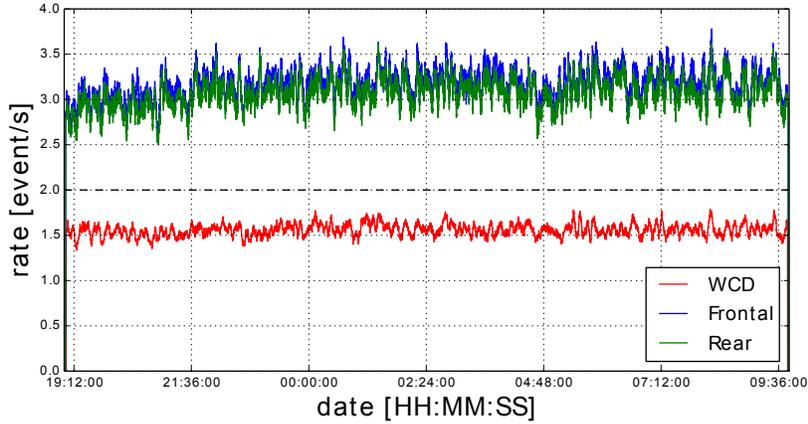}
\caption{In-coincidence event rate detected by the front (blue), rear (green) and the WCD (red). The WCD rate is $50\%$ lower than the detected by the panels due to due to its smaller acceptance angle than that of the hodoscope. The dashed line indicates the expected WCD rate taking into account the ratio between the angular apertures $\theta_H$ and $\theta_C$.}
\label{fig:RateWCDH}
\end{figure}

The aperture angle of the hodoscope $\theta_H$ at an inter-panel distance of $2.5$~m is around $50^{\circ}$, and the aperture of the whole detector (WCD + hodoscope) $\theta_C$ is roughly $32^{\circ}$. This means that several trajectories with high inclination angle will not be detected by the WCD, identifying only $\sim 62\%$ of the hodoscope events. In figure \ref{fig:RateWCDH}, we show the coincidence rate detected by the hodoscope planes and the WCD during $14$~hours. The mean rate for the panels is around $3.2$~events/s and for the WCD we have just $1.5$~events/s, which is  $\approx 50\%$ the rate of the hodoscope and it is lower than expected ($\sim 2$~events/s) due to the detection efficiency.

\begin{figure}[htb]
\centering
\includegraphics[width=0.48\columnwidth]{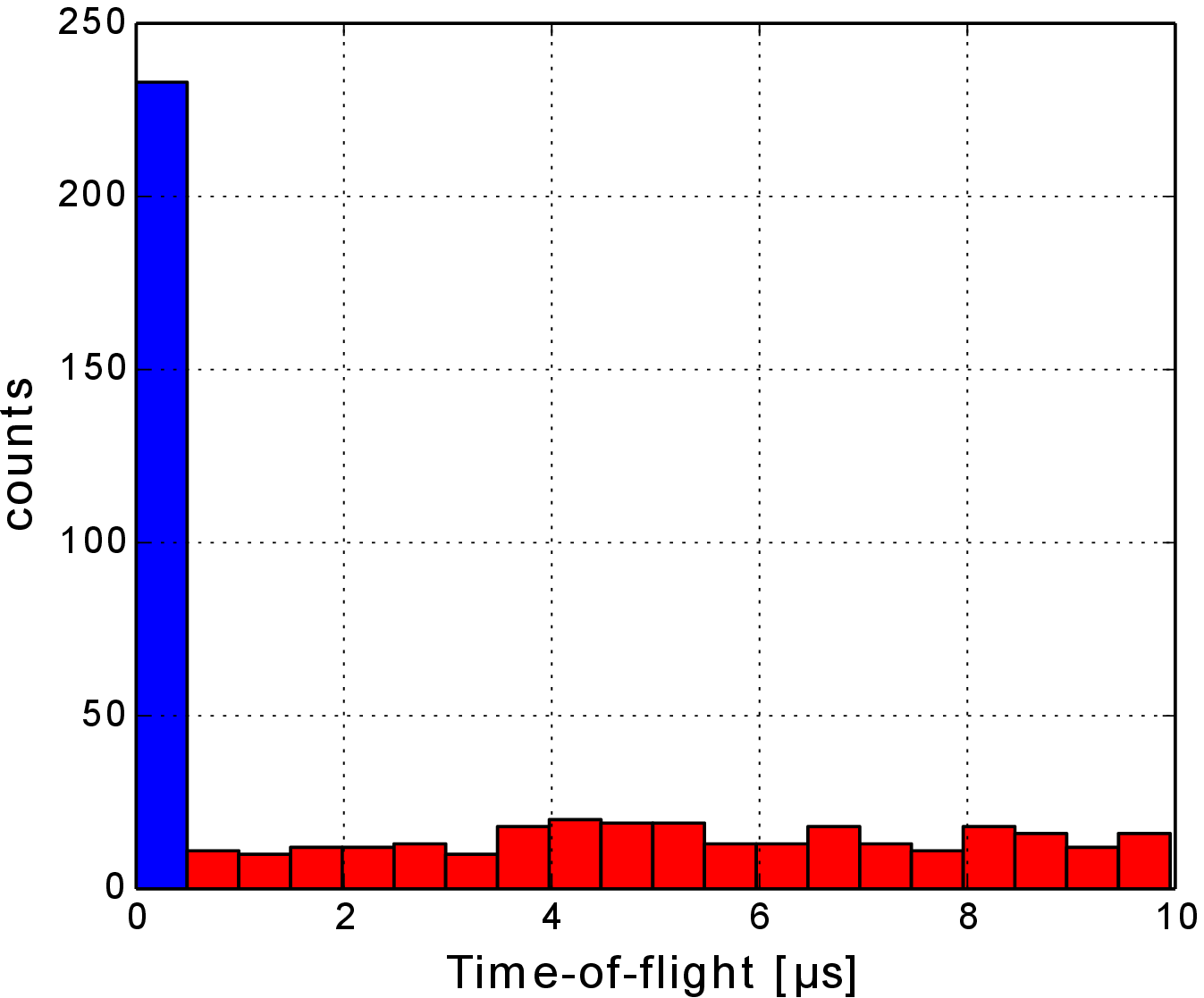}
\includegraphics[width=0.48\columnwidth]{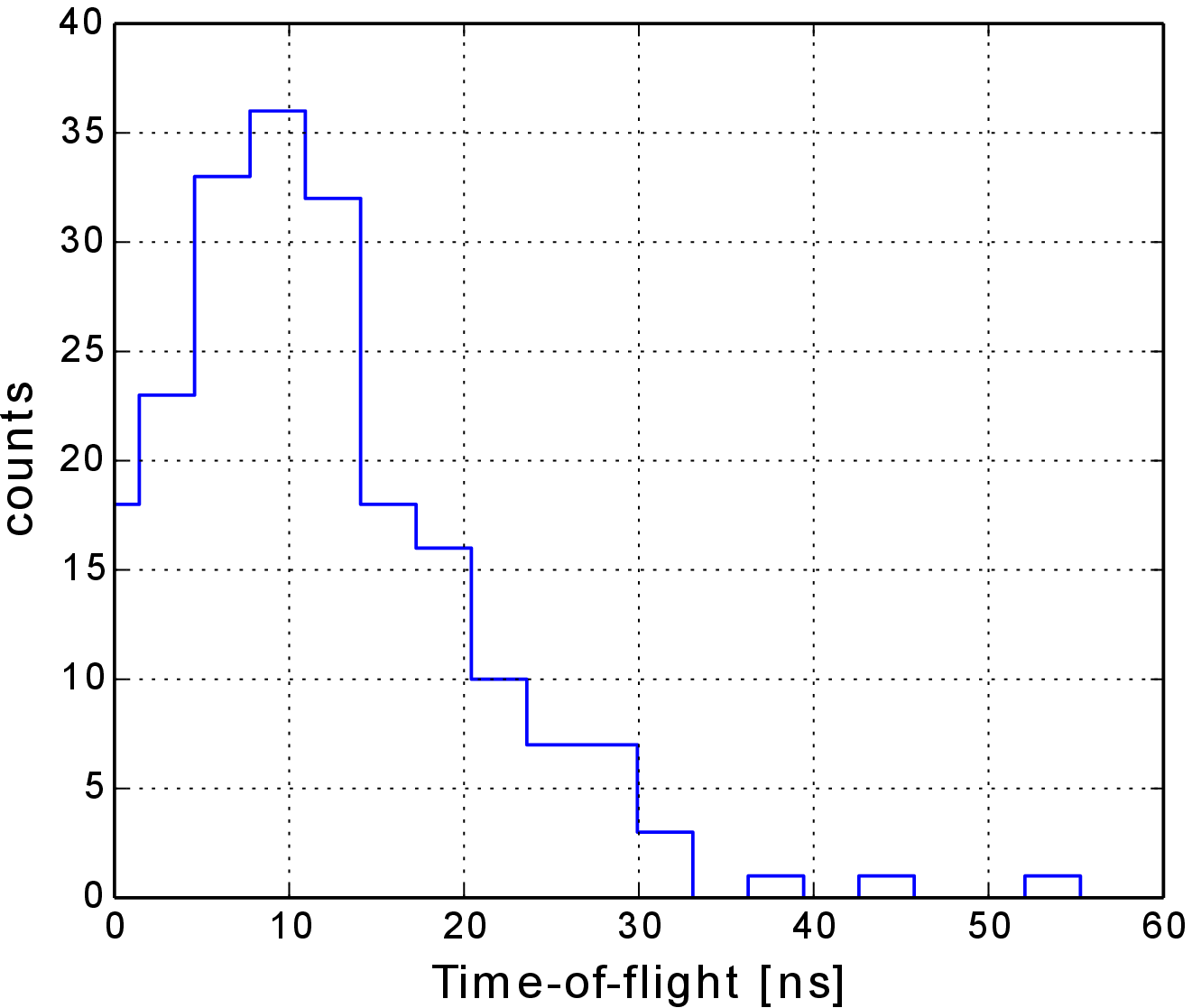}
\caption{Preliminary ToF measurements. The left side shows the ToF of single crossing particles (blue) and the time difference of particles impinging individually each panel (red). The right side shows the ToF distribution (mean $\sim$9.3\,ns) for single crossing particles.}
\label{fig:ToF}
\end{figure}

In figure \ref{fig:ToF}, we present preliminary results of ToF measurements. The left plate shows the ToF of single particles crossing the hodoscope (blue) and the time difference between two particles impinging individually each panel (red). The probability that two particles impinge individually each hodoscope panel creating a ToF signal like a single particle is negligible ($\sim~0.05\%$) under $200$~ns. The right side shows ToF details for single particles. The mean ToF ($\sim9.3$~ns) coincides with the expected range ($8.3$-$11.6$~ns) whose limits were defined as the ToF of a relativistic particle crossing the shortest ($2.5$~m) and the largest ($3.5$~m) hodoscope path. The signal delay in the scintillator bars enlarges the ToF measured range ($2.53$-$20.9$~ns).

\section{Some final remarks}
\label{conclusions}
In this paper, we presented the structural --mechanical and thermal-- simulations and the first calibration measurements of a hybrid muon telescope (scintillator hodoscope + WCD) designed to implement muography in the volcanoes of the Colombian Andes. Our instrument includes a hodoscope made by a pair of detection panels of plastic scintillator bars with an angular resolution of $32$~mrad for an inter-panel distance of $250$~cm. 

Furthermore, our design also incorporates particle identification techniques to filter the most common background noise sources in muography. A water Cherenkov detector allows to reduce noise signals coming from the soft-component of EAS (electrons and positrons), and multiple particle events using energy loss estimation. The WCD also detects fluctuations in the cosmic ray background at the observation place. Additionally, a picosecond Time-of-Flight system measures the direction and momentum of incident charged particles allowing the removal of backward and low momentum scattered muons. We estimate that MuTe can discriminate muons below  $0.4~\pm~0.1$~GeV/c taking into account the $138$~ps ToF resolution.

 The background noise due to the electromagnetic component of EAS was estimated to be $36\%$ of the collected data, while events corresponding to backward, forward and low momentum muons are $\sim 33.6\%$. As displayed in figure \ref{fig:WCDHod_rate}, the estimated multiple particle background was $30.4\%$ and the two-particle cases are the most probable in comparison with events involving several simultaneous particles. Such events release an average energy of $480$~MeV in the WCD.

The integrated flux recorded by the hodoscope pointing at $90^{\circ}$ with an aperture of $82^{\circ}$ drops drastically about two orders of magnitude compared to the total flux registered by the WCD at the observation point. Such flux reduction allows the multiple particle background to become more significant, decreasing the detector signal-to-background ratio.

After a complete analysis of the MuTe mechanical response corresponding to vibrations and tremors ranging from $1.6$~Hz to $7.5$~Hz, we found that MuTe structure would not undergo severe affectation. 

On the other hand, through the thermal simulations, we obtained that the maximum temperature in the hodoscope under the extreme Mach\'in volcano environmental conditions was $60^{\circ}$C at the rear side of the scintillation panels. An average wind speed of $30$~m/s generates a convection process in the front side of the panels, causing a temperature drop to $23^{\circ}$C. The WCD temperature, regulated by its water content, reaches at most 40$^{\circ}$C. This thermal analysis was considered for optimizing the SiPM operation \cite{PenaRodriguez2020}.

The angular resolution of MuTe ($32$~mrad) is similar to other experiments such as TOMUVOL ($8.7$~mrad) \cite{Crloganu2013}, MU-RAY ($15$~mrad) \cite{Ambrosino2014}, MURAVES ($8$~mrad) \cite{Cimmino2017}, and DIAPHANE ($100$~mrad) \cite{Lesparre2012}. Moreover, for the filtering of backwards muons, our ToF system has a better resolution ($\sim 138$~ps) compared to MURAVES ($400$~ps) and DIAPHANE ($1$~ns) \cite{ jourde2013experimental}. Our instrument incorporates particle identification techniques based on energy loss and momentum measurements to remove the background noise caused by low momentum muons, multiple particle events and electron/positrons.

\acknowledgments
We gratefully acknowledge the observations, suggestions and criticisms for the anonymous referees, improving the precision, presentation and clarity of the present work. The authors express our gratitude for the  financial support of  Departamento Administrativo de Ciencia, Tecnolog\'{\i}a e Innovaci\'on of Colombia (ColCiencias) under contract FP44842-082-2015 and to the Programa de Cooperaci\'on Nivel II (PCB-II) MINCYT-CONICET-COLCIENCIAS 2015, under project CO/15/02.  We are also very gratefull to LAGO and to the Pierre Auger Collaboration for their continuous support.  The simulations in this work were partially possible due to the computational support of the Red Iberoamericana de Computaci\'on de Altas Prestaciones (RICAP, 517RT0529), co-funded by the Programa Iberoamericano de Ciencia y Tecnolog\'{\i}a para el Desarrollo (CYTED) under its Thematic Networks Call. We also thank the permanent cooperation from the Universidad Industrial de Santander (SC3UIS) High Performance and Scientific Computing Centre. Finally, we would like to acknowledge the Vicerrector\'{\i}a Investigaci\'on y Extensi\'on Universidad Industrial de Santander for its permanent sponsorship. Finally, DSP would like to thank the School of Physics, the Grupo de Investigación en Relatividad y Gravitación, Grupo Halley and Vicerrector\'{\i}a Investigaci\'on y Extensi\'on of the Universidad Industrial de Santander for the warm hospitality during my post-doctoral fellowship.

\bibliographystyle{unsrt}
\bibliography{MuTe_bib.bib}

\appendix
\section{Estimation of MuTe power storage capacity}
\label{ApA}

The storage capacity required by the system in a day ($C_A$) is calculated using the modified Roger equation (\ref{cap_alma}) \cite{messenger2017photovoltaic}, i.e. 
\begin{equation}
C_a=\frac{E_c(1+F_s)}{\eta_{pb}\, \eta_{cdb} \, \eta_{rc} \, \eta_{pc}\, D_{b}} \, , \label{cap_alma}
\end{equation}
where $E_c$ is the load energy considering DC/DC converters, $F_s$ the scaling factor, $\eta_{pb}$ the efficiency of conductors, $\eta_{cdb}$ the efficiency of batteries, $\eta_{rc}$ the battery charge and discharge efficiency, $\eta_{pc}$ the charge controller efficiency, and $D_{b}$ corresponds to the battery depth of discharge. $E_c$ defined as follows
\begin{equation}
E_c=E_x+\frac{E_{\gamma}}{\eta_{dcdc}} \, ,
\label{two}
\end{equation}
where $E_{x}$ is the load energy which does not require DC/DC converters, $E_{\gamma}$ the load energy requiring DC/DC converters and, $\eta_{dcdc}$ the efficiency of the DC/DC converters. From (\ref{two}) with $E_{x}=58.57$~Wh and $E_{\gamma}=746.24$~Wh, we have $E_c=844.08$~Wh. 

Thus, by using $\eta_{pb}=0.97$, $\eta_{cdb}=0.95$, $\eta_{rc}=0.95$, $\eta_{pc}=0.98$, $D_{b}=0.8$ and $F_s=0.2$ (20\% oversizing), we obtained the value of $C_a=1472.75$~Wh, while the total storage capacity with which the system will count can be obtained by means of
\begin{equation}
C_{at}=\frac{C_aD_a}{V_{nb}} \, ,
\end{equation}
here $D_a$ is the total number of autonomy days and $V_{nb}$, is the nominal voltage of the battery bank. In this case, we set six days of autonomy with a nominal voltage of $12$~V, resulting in a total storage capacity of $C_{at}=736.38$~Ah. According to the criteria and environmental conditions presented above, we found the need for four batteries of $205$~Ah with a weight of $65$~Kg per battery and a discharge depth of $80\%$.

\end{document}